\documentstyle[twoside,fleqn,espcrc2,epsfig]{article}

\newcommand{\AmS}{{\protect\the\textfont2
  A\kern-.1667em\lower.5ex\hbox{M}\kern-.125emS}}

\title{Kondo effect and spin filtering in triangular artificial atoms}

\begin{document}

\author{
Gergely Zar\'and,*
\address{Lyman Laboratory of Physics, Harvard University, Cambridge MA 02145, USA}
\address{Department of Theoretical Physics, Institute of Physics, 
Budapest University of Technology and Economics,
H-1521 Hungary}
Arne Brataas,$^{\rm a}$ \address{Department of Physics, Norwegian University of Science and Technology, N-7491 Trondheim, Norway}
and David Goldhaber-Gordon\address{Department of Physics and Geballe Laboratory
for Advanced Materials, Stanford University, Stanford CA 94305}
}

\begin{abstract}
We study strongly correlated states in triangular artificial
atoms. Symmetry-driven orbital degeneracy of the single particle
states can give rise to an $SU(4)$ Kondo state with entangled
orbital and spin degrees of freedom, and a characteristic phase
shift $\delta = \pi/4$. Upon application of a Zeeman field, a
purely orbital Kondo state is formed with somewhat smaller Kondo
temperature and a fully polarized current through the device. The
Kondo temperatures are in the measurable range. The triangular
atom also provides a tool to systematically study the
singlet-triplet transitions observed in recent
experiments~\cite{Delft,Kastner}.
\end{abstract}
\maketitle

{\em Introduction.---} Submicron boxes of electrons known as
``quantum dots'' or ``artificial atoms'' have proved a fruitful
playground for studying single-particle quantized states,
interaction, and spin~\cite{Ashoorireview,Cobdenshells}. The role
of {\em orbital degeneracy} in the formation of strongly
correlated states in artificial atoms has not received the
scrutiny it deserves, perhaps because orbital degeneracy is so
hard to control experimentally in these systems. Accidental
degeneracies may result in spin $S > 1/2$ states with interesting
physical  properties~\cite{Glazman}, and in artificial atoms with
almost degenerate states  a singlet-triplet transition may also
occur, giving rise to interesting non-monotonic behavior in the
temperature dependence of the
conductance~\cite{Glazman,Delft,Hofstetter}. However, a typical
artificial atom, unlike its namesakes, has no spatial symmetries
and hence no orbital degeneracies. Instead, its ladder of orbital
states is well-described by random matrix
theory~\cite{Alhassidreview,ExperimentsRMT} This lack of symmetry
seems intrinsic to structures studied by lateral transport through
a patterned 2-dimensional electron gas (2DEG), as electrons must
enter and exit from leads which break the symmetry of the
confinement potential. In contrast, structures studied by vertical
transport may have cylindrical symmetry --- such artificial atoms
exhibit beautiful shell structure \cite{TaruchaAA}, demonstrating
that geometry can control a level spectrum. Combining the
advantages of vertical structures (controlled level spectrum) and
lateral structures (controlled tunneling strength) would be very
desirable. Here we show that this goal can be achieved by
carefully choosing the shape of a lateral structure.

\begin{figure}
\indent{
\epsfxsize=7cm
\epsffile{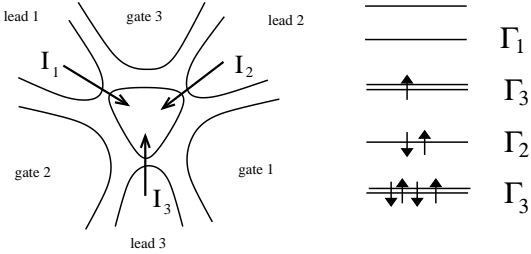}}
\caption{Proposed arrangement with triangular symmetry
and the structure of the four-fold  degenerate ground state of the
isolated TD.}
\label{fig:arrange}
\end{figure}

In a perfectly symmetrical  triangular artificial atom (TA)
each single particle state can be labeled
by its corresponding symmetry representation (see Fig.~\ref{fig:arrange}).
Generically about half the levels
are orbitally degenerate. To see this, let us imagine
slightly distorting a perfectly cylindrical box to a triangular
shape. Representation theory implies that states
with angular momentum $L_z  = \pm  3n$ split
into one-dimensional $\Gamma_1$ and $\Gamma_2$ representations,
while states with $L_z = \pm(3n + 1)$ or $L_z = \pm (3n - 1)$
remain degenerate, forming  two-dimensional  $\Gamma_3$
representations.  As a result, from $N$ originally degenerate states about
$2 N/3$ remain twofold degenerate, while the remaining $\sim N/3$ levels
split into $\sim 2 \times N/3$ non-degenerate levels by
the triangular symmetry.

As we discuss below, a singly-occupied $\Gamma_3$ state
gives rise to an $SU(N)$  Kondo state
with $N=4$, reflecting the fourfold degeneracy of the
ground state of the isolated TA (see Fig.~\ref{fig:arrange}).
In this state  orbital and spin fluctuations  are entirely entangled.
As opposed to the usual Kondo state with phase shift $\pi/2$,
this $SU(4)$ Kondo state is characterized by a phase shift
$\delta = \pi/4$, and is very similar
to states proposed recently for artificial molecules (two coupled artificial
atoms) \cite{Borda}.
The unusual phase shift could be detected  by integrating
the TA into an Aharanov-Bohm geometry \cite{Yacoby},
but could also be inferred from more standard transport measurents.
Application of a Zeeman field removes the spin degeneracy
of the ground state and drives the system to a purely
orbital Kondo state, in which the triangular atom acts as a spin
filter, transmitting only electrons with spin aligned to the external field.

The major difference between the artificial molecule system of
Ref.~\cite{Borda} and the present triangular arrangement is that
in the former case the orbital states of the full system are never
truly degenerate because of tunneling between the two sites, while
in a perfectly symmetrical TA the $\Gamma_3$ states are far closer
to degeneracy, split only by a small spin-orbit interaction.
Furthermore, the $\Gamma_3$ electronic wave functions on the TA
strongly overlap with each other, producing a large  Hund's rule
coupling, and hence a {\em triplet ground state} for  double
occupancy of the four-fold degenerate state.  Distorting the shape
of the TA with an external gate would gradually split the
degenerate $\Gamma_3$ states. Therefore, in this geometry one
could systematically study the  singlet-triplet transition without
applying any external magnetic field \cite{Delft,Kastner}.

{\em Model.---} First we focus on the charging
of a $\Gamma_3$ multiplet. At the Hartree-Fock level
the isolated TA can be described by:
\begin{eqnarray}
H_{\rm TA}  =  \sum_{\tau,\tau',\sigma}
d^\dagger_{\tau \sigma}
(E_{\tau \tau'} + \Delta E \;\delta_{\tau \tau'})
d_{\tau' \sigma}
-  J {\vec S}^{\;2}  \nonumber  \\
 +  {E_C \over 2} (n_+ + n_-)^2
+ {\tilde E_C\over 2} (n_+ - n_-)^2\;,
\label{eq:H_TA}
 \end{eqnarray}
where $d^\dagger_{\tau \sigma}$ creates an electron on the TA
within the $\Gamma_3$ multiplet
with spin $\sigma$ and orbital label $\tau = \pm$. The energy shift
$\Delta E$ is proportional to the (symmetrically-applied)
gate voltage and controls the charge on the atom,
while $E_{\tau \tau'}$ accounts for the splitting generated
by deviations from perfect triangular symmetry ($\sum_\tau E_{\tau \tau }=0$).
We denote the total number of electrons in state $\tau = \pm$ by
$n_\tau \equiv \sum_\sigma d^\dagger_{\tau \sigma}d_{\tau \sigma}$,
and ${\vec S} = {1\over 2} \sum_{\tau,\sigma,\sigma'}
d^\dagger_{\tau \sigma}{\vec \sigma}_{\sigma \sigma'} d_{\tau \sigma'}$
is the total spin of the atom. The terms proportional to $E_C$ and
${\tilde E_C}$ are generated by the Hartree interaction, while
that proportional to $J$ in  Eq.~\ref{eq:H_TA} is the
Hund's rule coupling, generated by exchange.

We describe the attached leads by
\begin{equation}
H_{\rm leads} = \sum_{\varepsilon,\sigma}
\sum_{j=1}^3 \epsilon \;a^\dagger_{\varepsilon \sigma j}
a_{\varepsilon \sigma j}
\end{equation}
where  $a^\dagger_{\varepsilon \sigma j}$
creates an electron in  lead $j$ with energy
$\varepsilon$ and spin $\sigma$,
and   $\{a^\dagger_{\varepsilon \sigma j},
a_{\varepsilon' \sigma' j'}\}= \delta_{j j'}
\delta_{\sigma\sigma'} \delta(\varepsilon-\varepsilon')$.

For a symmetrical TA we may usefully introduce a new conduction
electron basis through the unitary transformation,
$a_{m} \equiv \sum U_{m j } a_j$, with
$U_{mj} = e^{i 2\pi m\;j/3}/\sqrt{3}$.  States with
$\tau \equiv m=\pm 1$ transform
as $\Gamma_3$, while $a_{m=0}$
transforms as $\Gamma_1$, and cannot hybridize
with the $\Gamma_3$ doublet on the atom. Therefore, in this new
basis, for a perfectly symmetrical
TA  the hybridization between the atom and the leads
takes on a particularly simple form:
\begin{equation}
H_{\rm hyb} = V \sum_{\tau, \sigma} ( d^\dagger_{\tau \sigma}
\psi_{\tau \sigma} + {\rm h.c.})\;,
\end{equation}
with $\psi_{\tau \sigma} \equiv \int d\varepsilon\;
a_{\varepsilon \sigma \tau}$.

\begin{figure}
\indent{
\epsfxsize=7cm
\epsffile{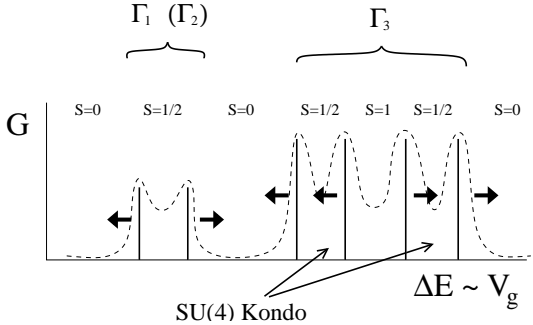}}
\caption{Structure of Coulomb blockade peaks as a function of
gate voltage, $V_g$. The total number of electrons on the dot 
increases by one at each Coulomb blockade peak with increasing 
$V_g$. The dashed line indicates the contribution from coherent
(Kondo) processes for a TA with spin $S\ne0$. Application of
a Zeeman field  shifts the peaks in directions indicated by 
the arrows.
}
\label{fig:Coulomb}
\end{figure}

{\em Coulomb blockade peaks.---}
We are most interested in the case of a four-fold degenerate
ground state arising from a singly-occupied $\Gamma_3$ level.
\footnote{Similar considerations would apply for the case of a single
hole on the $\Gamma_3$ multiplet.}  Such a state
can be identified experimentally by looking  at the height, magnetic field,
and temperature  dependence of the Coulomb blockade peaks.
In the regime where  the temperature $T$ is smaller than the
level spacing $\Delta$ of the dot, but is still larger than any
Kondo temperature or the tunneling rate $\sim 2\pi V^2$ to the leads,
the conductance around the Coulomb blockade peaks will
be dominated by sequential
tunneling. We use a standard rate  equation  formalism \cite{rate_equation}
to compute the linear conductance of the TD in this regime.
For a symmetrical
system the currents $I_j$ between leads $j$ and the atom are related to the
voltages $V_j$ applied on them by the conductance tensor,   $I_j =
\sum_{j'} G_{jj'} V_{j'}$, where $G_{j j'} =  {3\over 2}
G \delta_{j j'} - G/2$.    A schematic plot of the conductance
$G$ is shown in Fig.~\ref{fig:Coulomb}.  The four peaks associated with the
 $\Gamma_3$ state are spaced regularly in gate voltage, and their height
turns out to be numerically almost identical: For the first and
last peak we find $G_{1,\rm max}^{\Gamma^3} =  G_{4,\rm
max}^{\Gamma^3} = 0.9309{2e^2\over h} {\Gamma \over T}$, while the
second and third have a height $G_{2,\rm max}^{\Gamma^3} =  G_{3,
\rm max}^{\Gamma^3} = 0.9019 {2e^2\over h} {\Gamma \over T}$, with
$\Gamma = 2\pi V^2/3 \hbar$ the single particle tunneling rate
between the atom and one of the leads. Applying a Zeeman field to
the atom ({\em i.e.}, a field parallel to the underlying
2-dimensional electron gas), shifts the positions of the peaks in
the directions indicated by the arrows in Fig.~\ref{fig:Coulomb}
and also somewhat decreases their height. The $\Gamma_3$  peaks
are further distinguished by the  temperature dependence of the
conductance {\em between} them, since at low temperatures this
valley conductance increases for degenerate ground states due to
coherent Kondo-type correlations (see Fig.~\ref{fig:Coulomb})
\cite{Kondo_orig}.

{\em The {\rm SU(4)} Kondo state.---}
From now on we shall focus on the four-fold degenerate
ground state with $n_+ + n_- = 1$, expanded in states
$|\mu\rangle \equiv | \tau\;\sigma\rangle$.
When coupled to leads, this state exhibits a novel and interesting
Kondo effect. We compute the effective interaction between the leads and
the atom by integrating out virtual fluctuations to the
$ n_+ + n_- = 0$ and $n_+ + n_- = 2$ states, giving:
\begin{equation}
H_{\rm int} = \sum  J^{\mu\nu}_{\alpha\beta}
\; |\mu \rangle \langle \nu| \;
\psi^\dagger_\alpha\psi_\beta \;,
\end{equation}
where $\mu$, $\nu$, $\alpha$, and $\beta$ stand for the possible
combinations  of orbital and spin indices, $\alpha,\dots, \nu \in
\{\sigma,\tau\}$. The couplings  $J^{\mu\nu}_{\alpha\beta}$
are typically of order $V^2/ {E}_C$ and their explicit expression is
rather complicated. Fortunately, this 'bare' Hamiltonian
simplifies considerably upon scaling --- a renormalization
group analysis reveals that at  small energy scales (temperatures)
 the TA is simply described  by the  effective exchange Hamiltonian
\cite{Coqblin}:
\begin{equation}
H_{\rm eff}(T\to0) = \tilde J \sum_{\alpha,\beta = 1,..,4}
|\beta\rangle \langle \alpha|\; \psi_\alpha^\dagger  \psi_\beta \;.
\label{eq:H_eff}
\end{equation}

For a TA with perfect symmetry, this
 Hamiltonian produces a strongly correlated Kondo state
with SU(4) symmetry and entangled spin and orbital degrees of
freedom. 
At $T=0$ temperature,
both orbital and  spin indices are conserved and 
the transmitted electrons acquire only 
a phase shift  $\delta_{m,\sigma}$. 
In this Kondo state all four phase shifts equal in zero external field, 
and following the Friedel sum rule satisfy $\delta_{m,\sigma}= \pi/4$ \cite{Nozieres}.
This unusual phase shift should be measurable experimentally in Aharanov-Bohm 
arrangements \cite{Yacoby}.
In case of a finite Zeeman field the phase shifts acqire 
a spin dependence, but they remain independent of 
the orbital indices, 
$\delta_{m=\pm1,\sigma} =\delta_\sigma$, 
though $\delta_\sigma$ are related through the 
Friedel sum rule as $\delta_\uparrow + \delta_\downarrow = \pi/2$.

Knowledge of the phase shifts allows us to
compute the linear DC conductance through the TA at $T=0$
using the Landauer-Buttiker formalism \cite{Glazman,Buttiker}.
In the ground state,  
%
%
the $m=0$ combination of the leads is totally decoupled, and therefore 
$\delta_{0} = 0$. 
We can thus construct the
scattering matrix $S^\sigma_{mm'}$ within the basis
$\{a_{m\sigma}\}$, and then express the  scattering matrix $S^\sigma_{ij}$ by just
rotating back to the original basis: $S^\sigma_{ij} = \sum_{m,m'}
U^\dagger_{im}S^\sigma_{mm'} U_{m'j}$. The conductance matrix can then be
expressed through $S^\sigma_{ij}$, and for a perfectly
symmetrical  geometry we find that
$G(T\to 0) = G_Q {4\over 9} \sum \sin^2 \delta_\sigma$
where  $G_Q = 2e^2/h$ is the quantum conductance. 

For general $J^{\mu\nu}_{\alpha\beta}$ the Kondo
temperature $T_K$ cannot be given in a closed form. If, however,
 the dominant charge fluctuations are to the $n_+ + n_- =0$
state, then $J^{\mu\nu}_{\alpha\beta}\approx J \delta^{\mu}_{\beta}
\delta^{\nu}_{\alpha}$, and in the leading logarithmic approximation
 $T_K \approx \Delta\; e^{-1/4J}$. Thus $T_K$ can be tuned
to the experimentally accessible regime by appropriately
increasing the lead-TA conductances and making an atom with
sufficiently large level spacing. A triangular atom with side
200~nm would be challenging but feasible to fabricate, and should
satisfy the level spacing requirement.

It is interesting to study the effect of a Zeeman field,
$H_{\rm TA} \to H_{\rm TA} - B S^z$.  Experimentally,
this corresponds to a field applied {\em parallel} to the
surface of a lateral artificial atom \cite{footnote}.
For $B > T_K$ spin  fluctuations  on the atom  are
suppressed. Orbital fluctuations remain allowed, and
in the spin channel parallel to $B$ a purely orbital Kondo state
is formed.  For $J^{\mu\nu}_{\alpha\beta}\approx J \delta^{\mu}_{\beta}
\delta^{\nu}_{\alpha}$ we find that the orbital Kondo
temperature is somewhat suppressed compared to that of the
$SU(4)$ Kondo state, $T_K(B\to \infty)  \approx  \Delta
e^{-1/2J}\sim T_K(B=0)^2/\Delta$.

In contrast to the SU(4) Kondo state, this orbital state is
characterized by phase shifts $\delta_\uparrow = \pi/2$ and
$\delta_\downarrow = 0$. Interestingly, at $T=0$ the conductance
$G$ is approximately field-independent, since $\delta_\uparrow +
\delta_\downarrow = \pi/2$ as a consequence of the Friedel sum
rule. Thus the conductance in this regime should increase with
decreasing temperature even for $T < B$ and should saturate below
$T_K(B\to \infty)$, a clear signature of a purely orbital Kondo
state.

The  {\em polarization}  of the transmitted current, on the other
hand, $P = \sin^2\delta_\uparrow - \sin^2\delta_\downarrow$, does
depend on $B$, and the current becomes almost fully polarized for
$B > T_K$ at $T=0$. Therefore the triangular atom in this regime
could be used as a perfectly {\em spin filter} with high
conductance.

Distorting the shape of the TA, or applying a small magnetic field
perpendicular to the surface, splits the
$\Gamma_3$ multiplet. As with application of a parallel field,
the SU(4) Kondo is destroyed, replaced by a standard
spin-1/2 Kondo state with reduced Kondo temperature.
The conductance tensor in this case becomes
more complicated, however, and cannot be parametrized as before.

{\em Stability.---} Finally, let us comment on the conditions under which
the SU(4) and orbital Kondo states can be observed. Clearly,
it is experimentally very important to have as perfect
triangular symmetry as possible. Departures from symmetry
have two major consequences: (a) The  couplings
$J^{\mu\nu}_{\alpha\beta}$ become anisotropic.
Fortunately, this has little importance, since  the
SU(4) state is stable in the renormalization group sense, {\em i.e.},
these differences can be neglected in the Kondo
regime. (b) The $\Gamma_3$ multiplet splits.
This effect is insignificant until the
splitting is larger than $T_K$, which can be a substantial fraction of
the average level spacing. We therefore
believe  that these new Kondo states
can indeed be experimentally observed in a triangular
artificial atoms.

{\em Summary.---}
In a perfectly  triangular artificial atom
many of the single particle orbital states are degenerate by
symmetry.  We have shown that such orbital degeneracy
can result in the formation of an $SU(4)$ symmetric Kondo
state with phase shifts $\delta = \pi/4$, or
a purely orbital Kondo state in the presence of a strong external
parallel magnetic field. The triangular atom acts as a
spin polarizer in this regime. We have also suggested using triangular
atoms to study systematically the singlet-triplet
transition described in Ref.~\cite{Delft} and apparently
observed in Ref.~\cite{Kastner}.

{\em Acknowledgments.---}
We are grateful to B.I.  Halperin and W. Hofstetter
for stimulating discussions.   This research has been
supported  by   NSF Grants Nos.
DMR-9985978  and Hungarian Grants No. OTKA F030041,  T038162, and N31769.


\begin{thebibliography}{99}
\bibitem{footnote}
A small  magnetic field perpendicular  to the surface
would split the orbital degeneracy $\Gamma_3$ multiplet too.
\end{thebibliography}
\end{document}